\documentclass[]{aastex631}

\usepackage[figuresright]{rotating}
\shorttitle{AASTeX v6.3.1 Sample article}
\shortauthors{Ding et al.}
\graphicspath{{./}{figures/}}

\begin{document}

\title{Detection of Semi-detached eclipsing Binaries from TESS}
\correspondingauthor{KaiFan Ji}
\email{jkf@ynao.ac.cn}

\author[0000-0002-2427-161X]{Xu Ding}
\affiliation{Yunnan Observatories, Chinese Academy of Sciences (CAS), P.O. Box 110, 650216 Kunming, P. R. China}
\affiliation{Key Laboratory of the Structure and Evolution of Celestial Objects, Chinese Academy of Sciences, P. O. Box 110, 650216 Kunming, P. R. China}
\affiliation{Center for Astronomical Mega-Science, Chinese Academy of Sciences, 20A Datun Road, Chaoyang District, Beijing, 100012, P. R. China}

\author{KaiFan Ji}
\affiliation{Yunnan Observatories, Chinese Academy of Sciences (CAS), P.O. Box 110, 650216 Kunming, P. R. China}
\affiliation{Key Laboratory of the Structure and Evolution of Celestial Objects, Chinese Academy of Sciences, P. O. Box 110, 650216 Kunming, P. R. China}
\affiliation{Center for Astronomical Mega-Science, Chinese Academy of Sciences, 20A Datun Road, Chaoyang District, Beijing, 100012, P. R. China}
\affiliation{University of the Chinese Academy of Sciences, Yuquan Road 19\#, Shijingshan Block, 100049 Beijing, P.R. China}

\author{QiYuan Cheng}
\affiliation{Yunnan Observatories, Chinese Academy of Sciences (CAS), P.O. Box 110, 650216 Kunming, P. R. China}
\affiliation{Key Laboratory of the Structure and Evolution of Celestial Objects, Chinese Academy of Sciences, P. O. Box 110, 650216 Kunming, P. R. China}
\affiliation{Center for Astronomical Mega-Science, Chinese Academy of Sciences, 20A Datun Road, Chaoyang District, Beijing, 100012, P. R. China}
\affiliation{University of the Chinese Academy of Sciences, Yuquan Road 19\#, Shijingshan Block, 100049 Beijing, P.R. China}

\author{ZhiMing Song}
\affiliation{School of Information, Yunnan University of Finance and Economics, Kunming, China.}
\affiliation{Yunnan Key Laboratory of Service Computing, Kunming, China.}

\author{JinLiang Wang}
\affiliation{Yunnan Observatories, Chinese Academy of Sciences (CAS), P.O. Box 110, 650216 Kunming, P. R. China}
\affiliation{Key Laboratory of the Structure and Evolution of Celestial Objects, Chinese Academy of Sciences, P. O. Box 110, 650216 Kunming, P. R. China}
\affiliation{Center for Astronomical Mega-Science, Chinese Academy of Sciences, 20A Datun Road, Chaoyang District, Beijing, 100012, P. R. China}
\affiliation{University of the Chinese Academy of Sciences, Yuquan Road 19\#, Shijingshan Block, 100049 Beijing, P.R. China}

\author{XueFen Tian}
\affiliation{Communication And Information Engneering College, Yunnan Open University, 650500 Kunming, P. R. China}

\author{ChuanJun Wang}
\affiliation{Yunnan Observatories, Chinese Academy of Sciences (CAS), P.O. Box 110, 650216 Kunming, P. R. China}
\affiliation{Key Laboratory of the Structure and Evolution of Celestial Objects, Chinese Academy of Sciences, P. O. Box 110, 650216 Kunming, P. R. China}
\affiliation{Center for Astronomical Mega-Science, Chinese Academy of Sciences, 20A Datun Road, Chaoyang District, Beijing, 100012, P. R. China}
\affiliation{University of the Chinese Academy of Sciences, Yuquan Road 19\#, Shijingshan Block, 100049 Beijing, P.R. China}




\begin{abstract}
Semi-detached binaries, distinguished by their mass transfer phase, play a crucial role in elucidating the physics of mass transfer within interacting binary systems. To identify these systems in Eclipsing binary (EB) light curves provided by large-scale time-domain surveys, we have developed a methodology by training two distinct models that establish a mapping relationship between the parameters (orbital parameters and physical parameters) of semi-detached binaries and their corresponding light curves. The first model corresponds to scenarios where the more massive star fills its Roche lobe, while the second model addresses situations where the less massive star does so. In consideration of the O'Connell effect observed in the light curves, we integrated a cool spot parameter into our models, thereby enhancing their applicability to fit light curves that exhibit this phenomenon. Our two-model framework was then harmonized with the Markov Chain Monte Carlo (MCMC) algorithm, enabling precise and efficient light curve fitting and parameter estimation. Leveraging 2-minute cadence data from the initial 67 sectors of TESS, we successfully identified 327 systems where the less massive component fills its Roche lobe, alongside 3 systems where the more massive component fills its Roche lobe. Additionally, we offer comprehensive fundamental parameters for these binary systems, including orbital inclination, relative radius, mass ratio, and effective temperature.

\end{abstract}

\keywords{Binary stars (154) --- Eclipsing binary stars(444) --- Semi-detached binary stars(1443)}


\section{Introduction} 
Eclipsing binary systems are classified into three distinct categories based on their Roche lobe configurations: detached, semi-detached, and contact binaries \citep{Kopal+et+al+1959,Lucy+et+al+1968a,Lucy+et+al+1968b}. The interactions between binary systems lead to the formation of various fascinating objects, such as compact binaries, supernovae, gamma-ray bursts, X-ray binaries, pulsars, and cataclysmic variables \citep{Sana+et+al+2012}. In a semi-detached binary system, one component fills its Roche lobe and transfers mass to its companion. This interaction between the binary components plays a crucial role in stellar evolution and provides a valuable tool for validating related theoretical models \citep{Ibano+et+al+2006}. As a result, semi-detached binaries are of crucial importance in the study of various astrophysical phenomena, including mass transfer, accretion processes between components, and the loss of angular momentum via magnetic stellar winds.

Algol-type binaries are typically semi-detached systems, consisting of a main-sequence primary star ranging from mid-B to mid-F spectral types and an F-G-K type giant or subgiant secondary that either fills or overflows its Roche lobe \citep{Ibano+et+al+2006, Budding+et+al+1989}. In the formation of an Algol-type system, it is hypothesized that the massive star in a binary pair will exceed its Roche lobe capacity first, subsequently the more massive one is acting as a donor. This exchange subsequently reverses the mass ratio and accelerates the evolutionary progression of the secondary component \citep{Li+et+al+2022, Pols+et+al+1994}. Abundance of light curve data from surveys has facilitated the detection of semi-detached eclipsing binaries. \citet{Budding+et+al+2004} compiled a catalog of 411 Algol-type (semi-detached) binary systems. \citet{Paczy+et+al+2006} documented 2949 semi-detached Eclipsing Binaries within the All Sky Automated Survey catalog. \citet{Papageorgiou+et+al+2018} enumerated 4680 Northern Eclipsing Binaries exhibiting Algol-type light-curve Morphology in the Catalina Sky Surveys \citep{Marsh+et+al+2017}. Fitting light curves of Algol-type systems can provide crucial parameters for semi-detached binaries. These parameters include  the mass ratio, orbital inclination, temperature ratio, relative radii of the primary and secondary stars. Such parameters can reveal key evolutionary stages, such as whether it is the more massive or less massive star that fills its Roche lobe. \citet{Surkova+et+al+2004} compiled a sample of approximately 232 semi-detached binaries with stellar parameters. \citet{Lei+et+al+2024} obtained parameters for eight semi-detached eclipsing binaries from the Transiting Exoplanet Survey Satellite (TESS) survey \citep{Ricker+et+al+2015}. \citet{Li+et+al+2024} derived parameters for four targets from the Optical Gravitational Lensing Experiment project (OGLE), one of which features the more massive component filling its Roche lobe.

The Kepler \citep{Borucki+et+al+2010, Koch+et+al+2010} and TESS missions have provided high-precision light curves. \citet{Kirk+et+al+2016} found 2,878 eclipsing and ellipsoidal binary systems in the Kepler field of view. \citet{prsa+et+al+2022} identified 4,584 eclipsing binaries in TESS short-cadence observations of sectors 1 - 26. To precisely search for semi-detached binaries where either the more massive star or the less massive star fills its Roche lobe, we conducted a detailed modeling analysis of the EB-type light curves released by \citet{Gao+et+al+2025}. The Wilson–Devinney program \citep{Wilson+et+al+1990,Wilson+et+al+2012,Wilson+et+al+1971,Wilson+et+al+2010,Van+et+al+2007} and the Phoebe program \citep{prisa+et+al+2016}, though highly accurate, are computationally intensive, often requiring hours to days to process a single target. Consequently, efficiently extracting parameters from a large volume of light curves remains a formidable challenge in the field. Machine learning techniques have demonstrated potential in expediting parameter acquisition for binaries \citep{prsa+et+al+2008, Ding+et+al+2022, Xiong+et+al+2024}. We establish a mapping from parameters, including cool spot parameters, to light curves. Incorporating spot parameters primarily enhances the model's robustness, allowing for the fitting of light curves exhibiting the O'Connell effect \citep{Connell+et+al+1951, Milone+et+al+1968}. We train two neural network (NN) models: one for systems where the more massive component fills its Roche lobe, and another for systems where the less massive component fills its Roche lobe. We then utilize the Markov Chain Monte Carlo (MCMC) algorithm from the emcee program \citep{Foreman+et+al+2019}, which is implemented in Python, to effectively derive the posterior distribution of the parameters. Ultimately, 327 semi-detached binaries where the less massive star fills its Roche lobe, and three instances where the more massive star fills its Roche lobe, were identified. Additionally, we provide parameters for these semi-detached binaries, including orbital inclination, relative radius, mass ratio, and effective temperature. The structure of this paper is organized as follows: Section 2 describes the sample selection process, Section 3 details the development of the neural network model, Section 4 presents the parameter derivation for EB-type light curves, and Section 5 includes a discussion and conclusion.

\section{TESS sample selection} 
Using high-cadence 2-minute interval data from the initial 67 sectors of the TESS mission, \citet{Gao+et+al+2025} identified a total of 72,505 periodic variable stars. Within their comprehensive star catalog, 404 EB-type light curves were noted. After filtering out 66 candidates with classification probabilities below 0.5, which were potentially misidentified as semi-detached eclipsing binaries, we focused on the remaining 338 targets. We used the Python package lightkurve \footnote{https://docs.lightkurve.org/} to download the data for 338 systems authored by SPOC with an exposure time of 120 seconds. Notably, these light curves display the O'Connell effect, characterized by differing brightness levels during the maxima between eclipses. Consequently, we folded only the first three cycles of each light curve. The period values were sourced from the catalog periods calculated by \citet{Gao+et+al+2025}. To convert the flux measurements obtained from TESS into magnitudes, we applied the specified transformation equations.

\begin{equation}
    mag_i = -2.5 \times log10(flux_i)  
\end{equation}
\begin{equation} 
    \Delta{mag_i} = mag_i - \frac{\sum_{i=1}^n mag_i}{n}
\end{equation}

where $flux_i$ is the flux obtained from TESS. $\Delta{mag_i}$ is the normalized value of the corresponding magnitude.

We employed Gaussian Process Regression (GPR) \footnote{github.com/dfm/george/} to precisely fit the light curves of the 338 targets. The maximum magnitudes during the first and second out-of-eclipse phases are denoted as $Max.I$ and $Max.II$, respectively. The outcomes of these fittings are comprehensively summarized, with particular emphasis placed on $\Delta mag = Max.I - Max.II$, as illustrated in Figure 1. The noise present in the TESS light curves, predominantly clustered around $\sim 0.0012$ magnitudes, has been previously characterized by \citet{Ding+et+al+2024}. Obtaining the effective temperatures of these targets through cross-referencing with Gaia DR2 data \citep{Gaia+et+al+2018}, the distribution is shown in the right panel of Figure 1. Given the low noise environment and the inherent magnetic activity characteristic of late-type spectral stars, the $\Delta mag$ value effectively underscores the asymmetry evident in the light curves.

\begin{figure}[!ht]
\begin{center}

\begin{minipage}{5.7cm}
	\includegraphics[width=5.7cm]{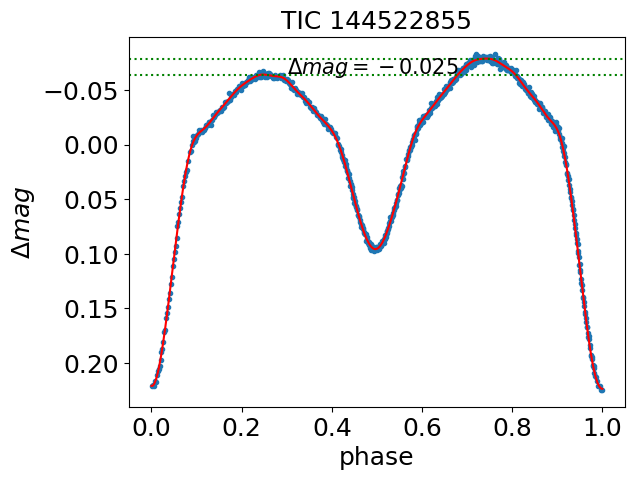}
\end{minipage}
\begin{minipage}{5.7cm}
	\includegraphics[width=5.7cm]{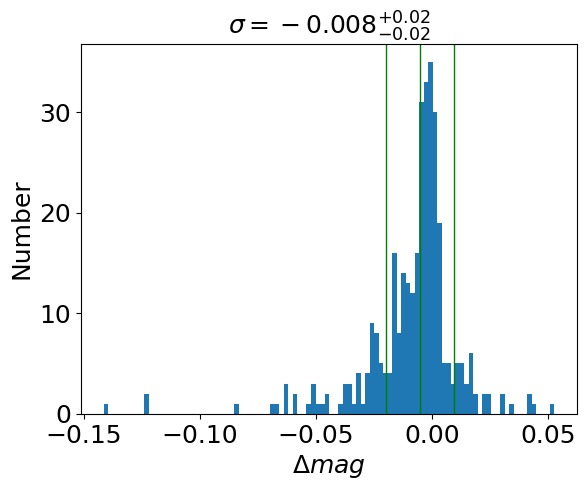}
\end{minipage}
\begin{minipage}{5.7cm}
	\includegraphics[width=5.7cm]{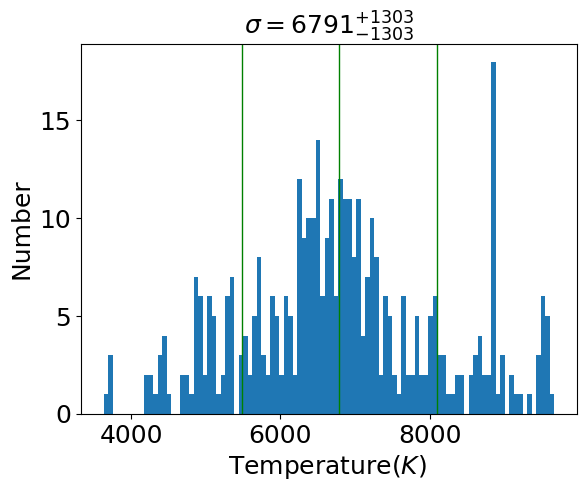}
\end{minipage}

\caption{Left: the blue dots represent the raw light curve data, while the red line indicates the fitting results. The green lines emphasize the Max.I and Max.II points within the light curves. Middle: the graph illustrates the $\Delta mag = Max.I - Max.II$ distribution specifically for 338 light curves. Right: the distribution of effective temperatures.
}\label{Fig4}
\end{center}
\end{figure}

\section{Building two neural network models} 
Semi-detached binary stars primarily involve two scenarios: one where the more massive component fills its Roche lobe, and another where the less massive component fills its Roche lobe. Based on these two situations, we have established training sample datasets and trained corresponding models to accurately capture the unique dynamics of each configuration.

\subsection{Two Sample data set}

In semi-detached binary systems, the light curves are significantly shaped by various parameters, including the mass ratio ($q$), orbital inclination ($incl$), effective temperatures of the primary ($T_1$) and secondary ($T_2$) stars, relative radii of the primary ($r_1=R_1/a$) and secondary ($r_2=R_2/a$) stars, with $a$ being the semi-major axis, gravity darkening coefficient ($g_1$ and $g_2$), bolometric albedos ($A_1$ and $A_2$), as well as the passband and properties of any star spots. Star spots are characterized by their colatitude ($colat$), longitude ($long$), angular radius ($radius$), and temperature ratio relative to the local intrinsic value ($relteff$). For the O'Connell effect, it is common practice to incorporate cool spots or hot spots into the fitting process \citep{Shi+et+al+2021}.

We first generate samples for the scenario where the less massive component fills its Roche lobe. To generate accurate light curves for semi-detached binaries, we employ the Phoebe program, which utilizes the fundamental parameters described in the previous section. To achieve the objective where the less massive component fills its Roche lobe, it is sufficient to set the relative radius of the primary star ($r_1$).
The relative radius of the primary star ($r_1$) is distributed within the interval [0.01, $\frac{R_{\rm RL}}{a}$], where the $R_{\rm RL}$ is the Roche lobe radius \citep{Eggleton+et+al+1983}.
The Roche lobe radius ($R_{\rm RL}$) when the the more massive component is filling its Roche lobe is estimated using the formulation provided by 
\begin{equation}
 \frac{R_{\rm RL}}{a} = \frac{0.49q^{-2/3}}{0.6q^{-2/3}+\ln(1+q^{-1/3})}
\end{equation}
The chosen passband corresponds to TESS:T, covering a broad wavelength spectrum from 600 to 1000 nm, as detailed by \citep{Ricker+et+al+2015}. The mass ratio ($q$) is uniformly spread across the [0,1] interval. The orbital inclination ($incl$) varies from 50 to 90 degrees. The primary star's temperature ($T_1$) is randomly selected from a range of 3500 K to 12500 K, while the secondary star's temperature ($T_2$) is determined within the bounds of 3500 K and up to 1.2 times $T_1$. \citet{Xiong+et+al+2024} set the effective temperature ratio ($T_2/T_1$) of the training samples in the range [0.2, 2.25], but the maximum temperature ratio of the 77 actual semi-detached binaries they analyzed was 0.989. Additionally, \citet{Ibano+et+al+2006} obtained parameters for 61 semi-detached binaries, with the maximum temperature ratio being 0.874. Therefore, a maximum value of 1.2 is chosen for $T_2/T_1$ in this study. Although this is not a highly precise value, it effectively covers the most common temperature ratios in the actual samples and allows for the generation of more valid samples within this range.

We place a cool spot on the primary star.
The cool spot's longitude ($long$) on the primary star is randomly positioned between 0 and 360 degrees, with its angular radius ($radius$) ranging from 0 to 35 degrees. The temperature ratio relative to the local intrinsic value ($relteff$) is set between 0.6 and 1. For stars with effective temperatures below 8000K, the gravity-darkening coefficient ($g$) is fixed at 0.32, and the bolometric albedo ($A$) at 0.5 \citep{Rafert+Twigg+1980}. Conversely, for stars with effective temperatures exceeding 8000K, these values are adjusted to 1 for both $g$ and $A$ \citep{Rafert+Twigg+1980}. 

We proceed to generate samples for the scenario where the more massive star fills its Roche lobe. Unlike the situation where the less massive star occupies its Roche lobe, here we adjust the relative radius of the secondary star ($r_2$) based on the Roche lobe radius ($R_{\rm RL}$) by employing the reciprocal of the mass ratio ($q^{-1}$) in equation 3. All other parameter ranges remain unchanged. Each of the two datasets comprises 300,000 samples.



\subsection{Two neural network models} 

This study employs a multilayer perceptron (MLP) neural network to model the relationship between input parameters and phase-magnitude light curves. For both datasets under investigation, an identical network architecture is implemented to ensure methodological consistency. The neural network configuration, as illustrated in Figure 2, consists of three primary components: an input layer, multiple hidden layers, and an output layer. The numerical annotations in Figure 2 indicate the node count for each respective layer.

The input layer is designed with nine nodes, corresponding to the following astrophysical parameters: $T_1, T_2, incl, q, r, relteff, colat, long, radius$. In the case where the more massive (primary star) component fills its Roche lobe, the input radius parameter ($r = r_2$) represents the relative radius of the less massive component (secondary star). Conversely, when the less massive star (secondary star) fills its Roche lobe, the input radius parameter ($r = r_1$) denotes the relative radius of the more massive component (primary star).

To effectively model non-linear relationships within the data, activation functions are implemented in both the input and hidden layers. The Rectified Linear Unit (ReLU) \citep{He+et+al+2015} activation function is selected for its demonstrated efficacy in handling complex patterns and its computational efficiency. Model performance is evaluated using the Mean Square Error (MSE) as the loss function, which quantifies the discrepancy between predicted and observed values. The Adam optimization \citep{Kingma+et+al+2014} 
 algorithm is employed for parameter tuning, leveraging its adaptive learning rate capabilities and computational efficiency. During model development, the validation set loss function trajectory is monitored to identify the optimal model configuration for subsequent training on the training dataset. The neural network architecture we have developed can effectively establish a mapping between parameters and light curves, thereby facilitating the rapid generation of light curves.
 
\begin{figure}[!ht]
\begin{center}

\begin{minipage}{16cm}
	\includegraphics[width=16cm]{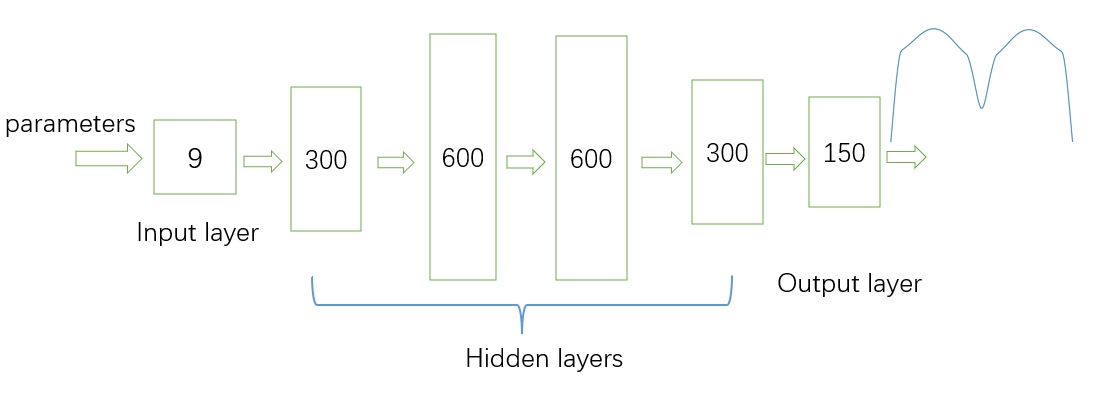}
\end{minipage}

\caption{This figure serves as a visual depiction of the neural network’s structural framework, wherein the numerical labels correspond to the neuron count within each designated layer.}\label{Fig1}
\end{center}
\end{figure}

\vspace{2em}
\subsection{Precision of light curves generated by two neural network models}
The original dataset was partitioned into three distinct subsets following an 8:1:1 distribution: training, validation, and test sets. The training set was employed to construct a neural network model, which underwent a learning process to optimize its parameters. Upon completion of the training phase, the final iteration of the model was assessed for its effectiveness using the validation set. Finally, the precision of the model's light curve generation was measured by evaluating its performance against the independent test set.

Initially, the model was tested under conditions where the less massive star fills its Roche lobe. The precision of the light curves generated by the model was evaluated by analyzing the standard deviation of the residuals. This evaluation involved a comparison of the residual distributions from synthetic light curves produced by Phoebe and those generated by the model. As illustrated in the left panel of Figure 3, the blue light curve was generated by Phoebe with the following input parameters: $T_1 = 4374$ K, $T_2 = 3852$ K, $incl = 80.06^{\circ}$, $q = 0.33$, $r_1 = 0.21$, $relteff = 0.90$, $radius=12.83^{\circ}$, $colat=120^{\circ}$, $long=-318^{\circ}$. The model, using the same parameters, produced the orange light curve in the same panel. The standard deviation of the residuals between these two curves was calculated to be 0.0009. Furthermore, the standard deviation of residuals was computed across a sample of 30,000 light curves. The right panel of Figure 3 shows that the standard deviations of the residuals are primarily distributed around $0.0006^{+0.0003}_{-0.0002}$. 

Next, the model was tested under conditions where the more massive star fills its Roche lobe. The precision of the light curves generated by the model was again evaluated by examining the standard deviation of the residuals. This assessment involved comparing the residual distributions from synthetic light curves produced by Phoebe and those generated by the model. As depicted in the left panel of Figure 4, the blue light curve was generated by Phoebe using the following input parameters: $T_1 = 4696$ K, $T_2 = 3726$ K, $incl = 80.54^{\circ}$, $q = 0.73$, $r_2 = 0.15$, $relteff = 0.93$, $radius=12.08^{\circ}$, $colat=74.08^{\circ}$, $long=-116^{\circ}$. The model, employing the same set of parameters, produced the orange light curve in the same panel. The standard deviation of the residuals between these two curves was calculated to be 0.0004. Additionally, the standard deviation of residuals across a sample of 30,000 light curves was computed. The right panel of Figure 4 illustrates that the standard deviations of the residuals are predominantly distributed around $0.0007^{+0.0006}_{-0.0002}$. Although the number of hidden layers and neurons in this work may not be optimal, they suffice to meet the requirements in terms of generating light curve accuracy.

\begin{figure}[!ht]
\begin{center}

\begin{minipage}{8cm}
	\includegraphics[width=8cm]{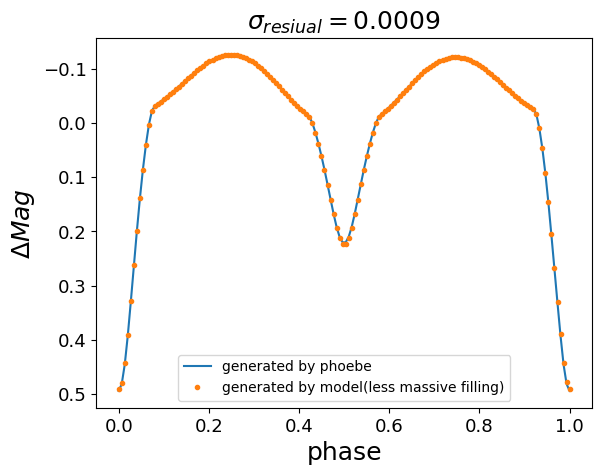}
\end{minipage}
\begin{minipage}{8cm}
	\includegraphics[width=8cm]{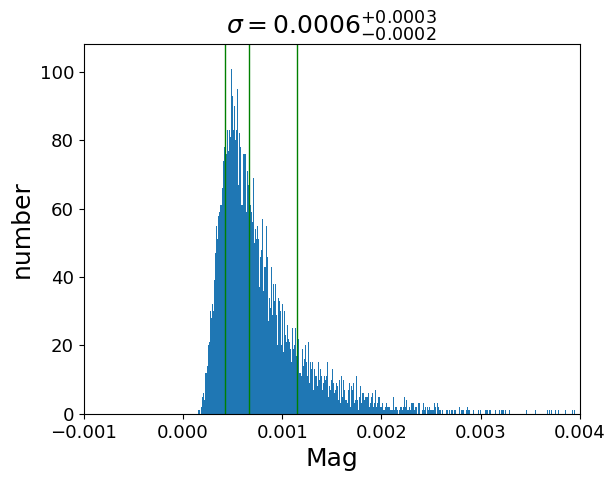}
\end{minipage}

\caption{The blue light curve on the left was synthesized by Phoebe, whereas the model (less massive filling), using the same input parameters, produced the orange curve. Meanwhile, the distribution shown on the right illustrates the standard deviation of the residuals between the model’s predicted light curves and Phoebe’s synthetic curves.}\label{Fig2}
\end{center}
\end{figure}

\begin{figure}[!ht]
\begin{center}

\begin{minipage}{8cm}
	\includegraphics[width=8cm]{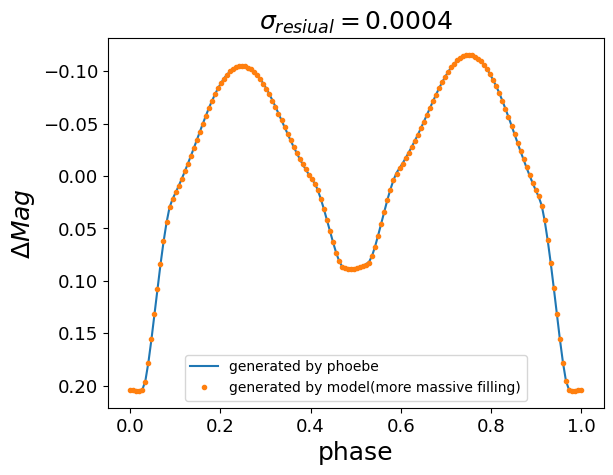}
\end{minipage}
\begin{minipage}{8cm}
	\includegraphics[width=8cm]{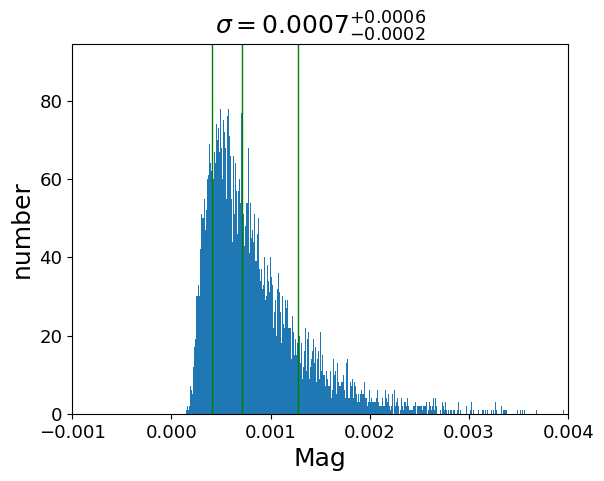}
\end{minipage}

\caption{On the left, the blue light curve represents data synthesized by Phoebe, while the orange curve illustrates the output generated by the model (more massive filling) when using the same input parameters. On the right, we present the distribution of the standard deviation for the residuals between the light curves predicted by our model and those synthetic curves created by Phoebe.}\label{Fig2_1}
\end{center}
\end{figure}

\vspace{4em}
\section{Solutions of EB-type light curves}
Using two neural network (NN) models that have been trained and fine-tuned, in combination with the Markov Chain Monte Carlo (MCMC) algorithm, we derive the posterior distributions of the parameters characterizing semi-detached binaries. Semi-detached binary systems where the more massive component fills its Roche lobe are less numerous because the evolution timescale of the more massive component is very short. Therefore, most semi-detached binary systems involve the less massive star filling its Roche lobe.

\subsection{Less massive star fills its Roche lobe: 327 targets}
We obtained 327 systems where the less massive star fills its Roche lobe and provided their fundamental parameters. By integrating a neural network model (specifically designed for systems where the less massive star fills its Roche lobe) with the Markov Chain Monte Carlo (MCMC) algorithm, we achieved highly accurate fits to the light curves of these systems. To exemplify our methodology, we present a detailed analysis of TIC 144522855 shown in the left panel in Figure 1, demonstrating the process of extracting its system parameters. Subsequently, we provide comprehensive fitting results for four additional notable targets, further validating the robustness and precision of our approach. This combination of advanced modeling techniques ensures reliable parameter estimation and enhances our understanding of such binary systems.

We commenced our analysis by applying the Neural Network (NN) model in tandem with the Markov Chain Monte Carlo (MCMC) algorithm to derive the posterior distribution of the parameters for the designated target, TIC 144522855. Our methodology involved deploying 50 parameter space walkers. To ascertain the convergence of the MCMC chain, we instituted a criteria to determine an optimal chain length, ranging from 10 to 20 times the integrated autocorrelation time, as recommended by \cite{Conroy+et+al+2020} and \cite{Li+et+al+2021}. The effective temperature of the primary star ($T_1$) was set as 6880 K by cross-referencing with the Gaia Data Release 2 (DR2) data \citep{Gaia+et+al+2018}. To ensure convergence, we extended the chain length to a value surpassing 20 times the integrated autocorrelation time that, as specified for the nine input parameters of the NN model, ranges between 18 and 185 steps. After conducting a thorough and comprehensive assessment, a chain length of 5000 steps was ultimately chosen. The light curve underwent 250,000 iterations of the MCMC parameter search post an initial burn-in phase of 5000 steps. Out of these, the final 2000 steps were retained for analysis, and a corner plot was subsequently generated to visualize the posterior distribution of the parameters ($T_2, incl, q, r_1, relteff, colat, long, radius$) as shown in Figure 5. In the fitting process, we also took into account slight adjustments in phase shift and magnitude ($ShiftPhase, ShiftMag$) in Figure 5. Figure 6 presents the original light curve as blue dots, with the red line representing the light curve reconstructed by our model using the derived parameters. The 'x' points on the same panel correspond to the light curve generated by Phoebe with identical parameters. The goodness-of-fit ($R^2$) between the Phoebe-generated light curve and the original light curve was found to be an impressive 0.998.

Building on this robust methodology, we applied the NN model (with the less massive star filling its Roche lobe) combined with the MCMC algorithm to fit the light curves of four prominent targets identified as TIC 434330408, TIC 434508647, TIC 34264274, and TIC 416024040, as illustrated in Figure 7. The fundamental parameters can be seen in Table 1.
Each of these fits yielded a goodness-of-fit ($R^2$) exceeding 0.98, underscoring the high precision and reliability of our fitting approach. A catalog \footnote{github.com/dingxu6207/PaperData/tree/main/Data1} provided by this work encapsulates the fundamental parameters for 327 systems where the less massive star fills its Roche lobe.

\begin{figure}[!ht]
\begin{center}
\begin{minipage}{18cm}
	\includegraphics[width=18cm]{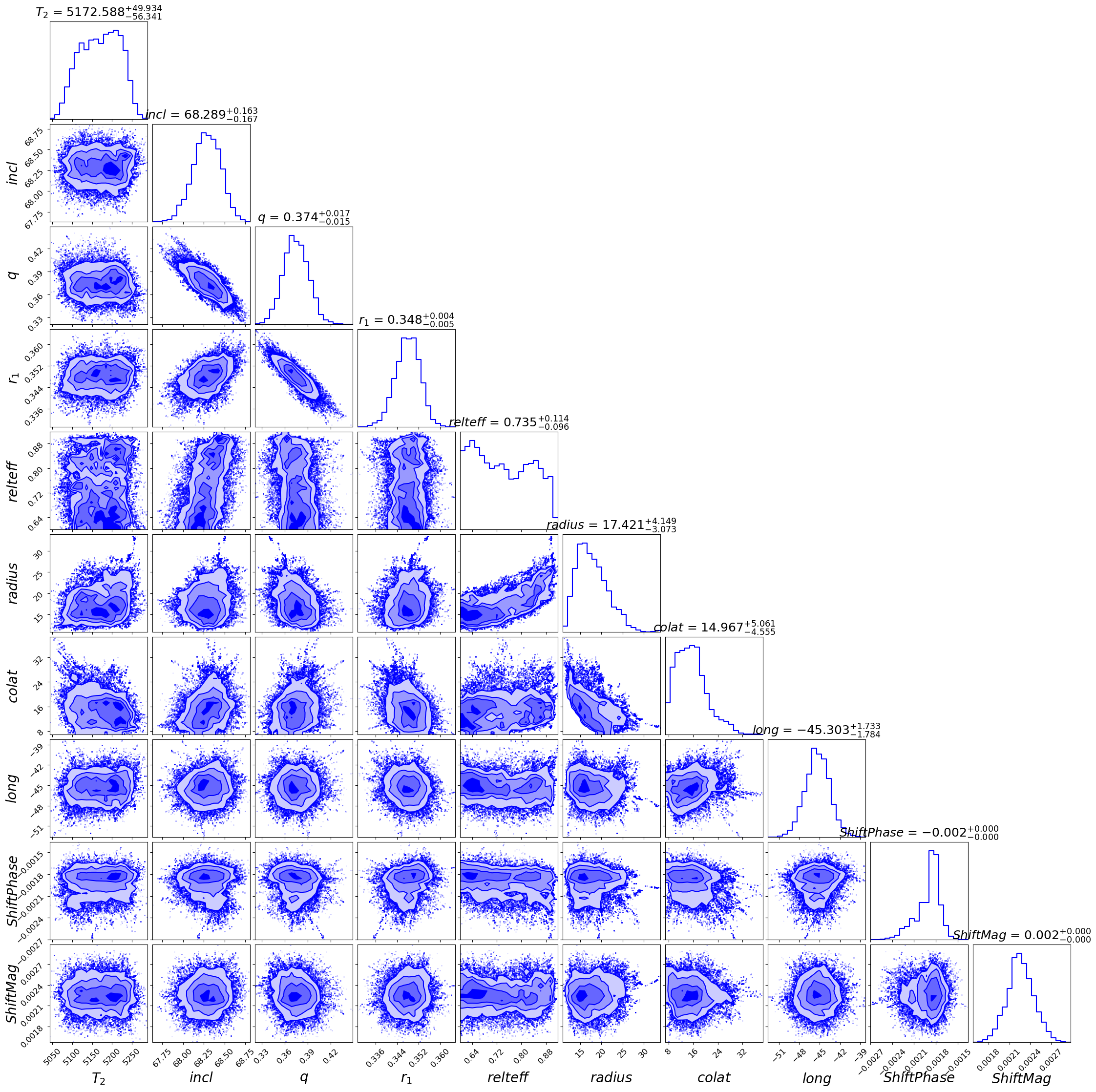}
\end{minipage}

\caption{The posterior parameter distributions for TIC 144522855 are presented.}\label{Fig5}
\end{center}
\end{figure}

\begin{figure}[!ht]
\begin{center}
\begin{minipage}{10cm}
	\includegraphics[width=10cm]{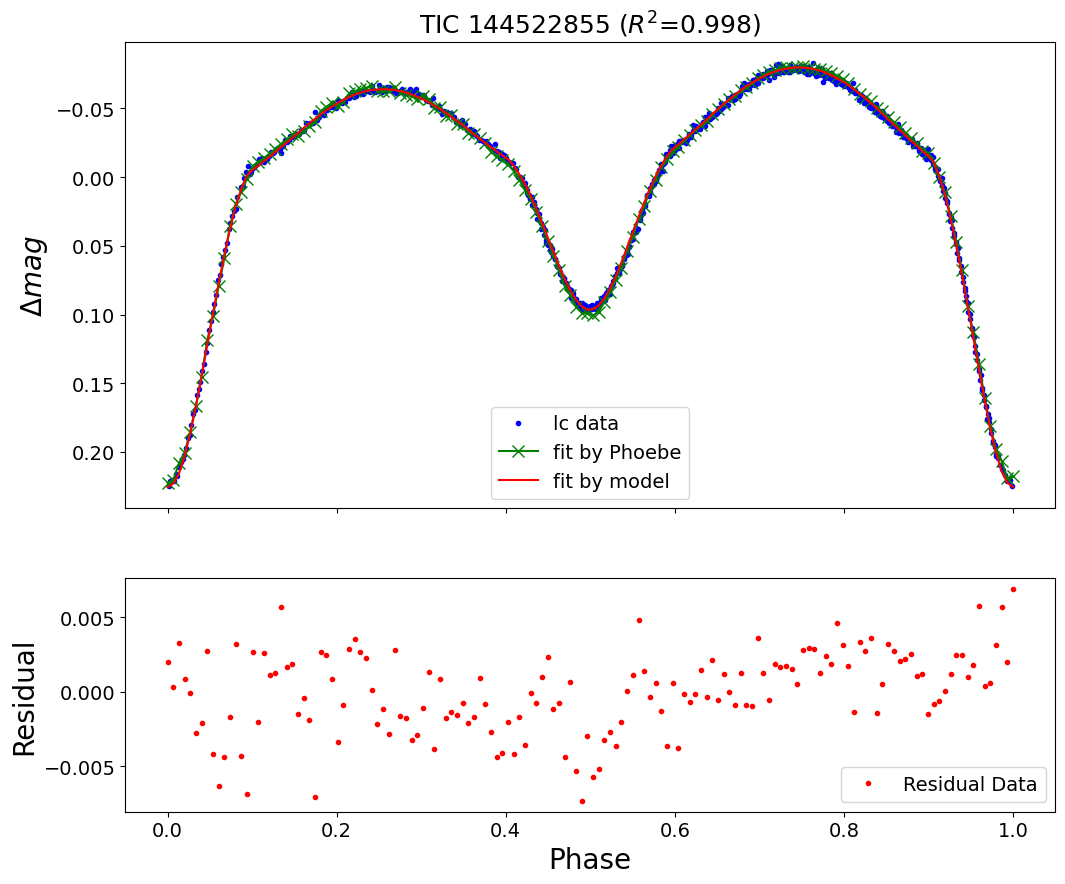}
\end{minipage}
\caption{Fitting Results for TIC 144522855 are presented.}\label{Fig6}
\end{center}
\end{figure}

\begin{figure}[!ht]
\begin{center}
\begin{minipage}{7.2cm}
	\includegraphics[width=7.2cm]{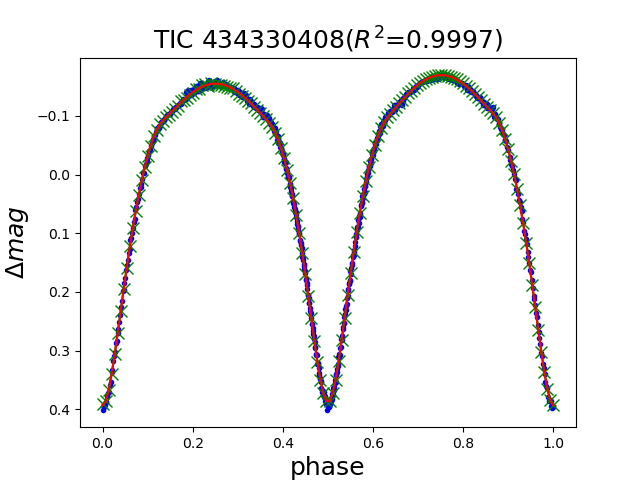}
\end{minipage}
\begin{minipage}{7.2cm}
	\includegraphics[width=7.2cm]{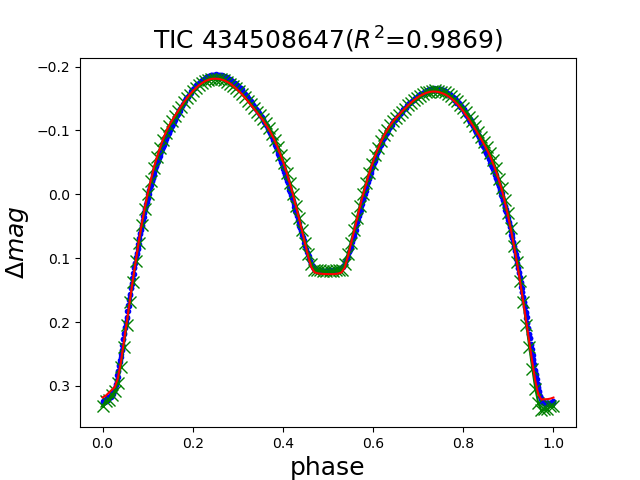}
\end{minipage}
\begin{minipage}{7.2cm}
	\includegraphics[width=7.2cm]{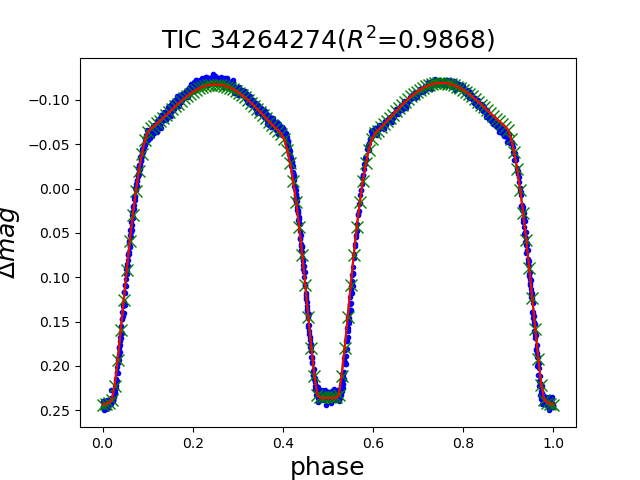}
\end{minipage}
\begin{minipage}{7.2cm}
	\includegraphics[width=7.2cm]{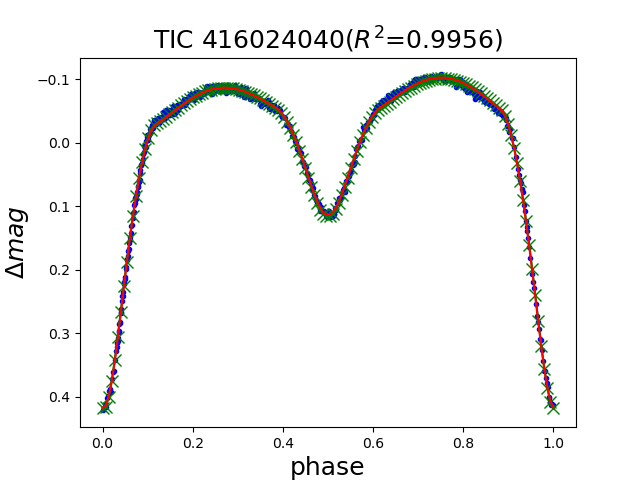}
\end{minipage}

\caption{The fitting results for four targets are depicted, with the blue dots indicating the original light curves. The red lines represent the light curves generated by the trained model (less massive filling), and the green 'x' dots denote the light curves produced by Phoebe.}\label{Fig7}
\end{center}
\end{figure}

\clearpage
{
\tiny
\begin{center}
\begin{longtable}{ccccccccccccccc}
\caption{Parameters of Seven Semi-Detached Binaries: First 4 (Less Massive Filler) and Last 3 (More Massive Filler)
}\label{Table 2_1}\\
\hline\hline                          
name     &$T_1$(K)   &$T_2$(K) &$incl(\circ)$  &$\sigma_{incl}$  &$q$  &$\sigma_{q}$  &$r_1$  &$r_2$  &$R^2$                   \\
\hline
\endhead
\hline

TIC 34264274 & 5926 & 6022   & 83.498 & 1.343 & 0.121 & 0.009 & 0.384 & 0.217 & 0.987 \\
TIC 416024040 & 6781 & 5226  & 77.311 & 1.377 & 0.287 & 0.028 & 0.388 & 0.277 & 0.996 \\
TIC 434330408 & 7109 & 7124  & 80.773 & 0.651 & 0.487 & 0.120 & 0.400 & 0.318 & 1.000 \\
TIC 434508647 & 7034 & 5983  & 84.452 & 2.083 & 0.290 & 0.002 & 0.488 & 0.278 & 0.987 \\
TIC 282050293 & 6317 & 4222  & 69.879 & 0.122 & 0.145 & 0.004 & 0.550 & 0.156 & 0.991 \\
TIC 32284829 & 5586 & 4998   & 89.147 & 0.571 & 0.559 & 0.061 & 0.432 & 0.326 & 0.994 \\
TIC 449517017 & 7378 & 6575  & 89.749 & 0.309 & 0.618 & 0.006 & 0.423 & 0.338 & 0.998 \\

\hline
\end{longtable}
\end{center}
}

\subsection{More massive star fills its Roche lobe: 3 targets}
For TIC 282050293, TIC 32284829, and TIC 449517017, the model assuming the less massive star fills its Roche lobe yielded unsatisfactory fitting results. We therefore adopted the more massive star filling model, which significantly enhanced the fitting performance, with goodness-of-fit ($R^2$) values surpassing 0.99 for all three systems. The corresponding parameters are listed in Table 1, confirming these targets as robust candidates for systems where the more massive star fills its Roche lobe. The fitting results are illustrated in Figure 8.

\begin{figure}[!ht]
\begin{center}
\begin{minipage}{5.7cm}
	\includegraphics[width=5.7cm]{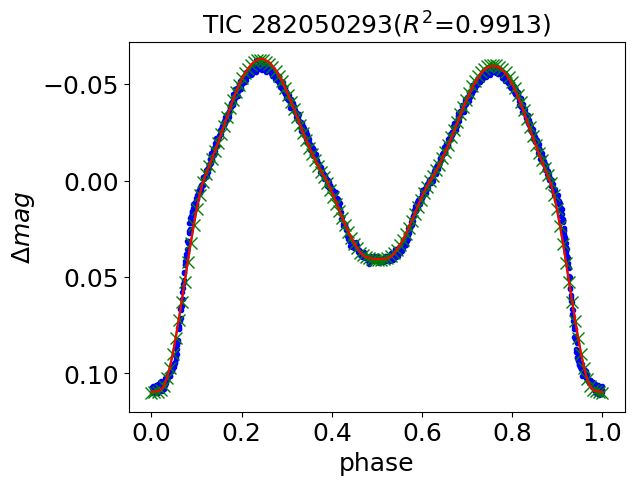}
\end{minipage}
\begin{minipage}{5.7cm}
	\includegraphics[width=5.7cm]{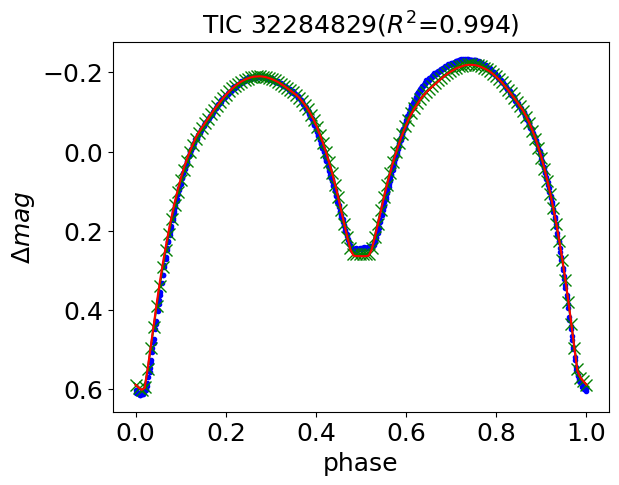}
\end{minipage}
\begin{minipage}{5.7cm}
	\includegraphics[width=5.7cm]{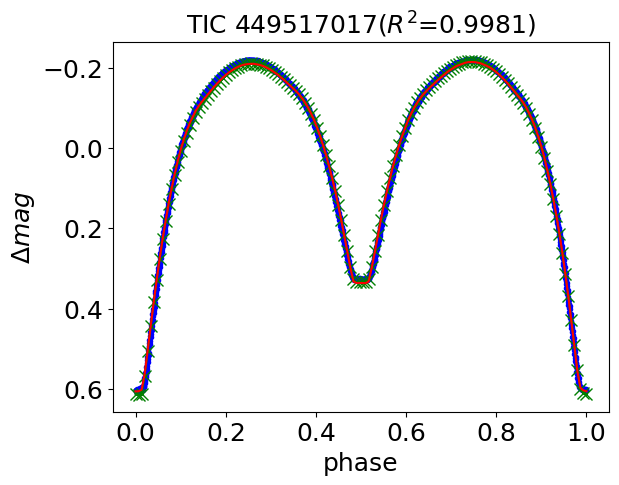}
\end{minipage}
\caption{The fitting results for three targets are depicted, with the blue dots indicating the original light curves. The red lines represent the light curves generated by the trained model (more massive filling), and the green 'x' dots denote the light curves produced by Phoebe.}\label{Fig8}
\end{center}
\end{figure}

\subsection{Contact binary: 5 targets}
The five targets TIC 189476500, TIC 197933923, TIC 220654235, TIC 254748090, and TIC 274645120 cannot be fitted using models where either the more massive star or the less massive star fills the Roche lobe. Building upon the work of \citet{Ding+et+al+2022}, who developed a model mapping the parameters of contact binary stars to their respective light curves, we incorporated spot parameters into the model. In the model we built, the input layer consists of 9 nodes, which correspond to the 9 input parameters ($T_1, q, incl, f, T_2/T_1, colat, long, radius, relteff$). Similar to the descriptions in Section 3.1, $T_1$, $q$, and $incl$ denote the effective temperature of the primary star, the mass ratio, and the orbital inclination respectively. The parameters ($colat$, $long$, $radius$, and $relteff$) are associated with the characteristics of the cool spot. What sets this apart is the incorporation of the fill-out factor ($f$) and the ratio of the secondary star's temperature to that of the primary star ($T_2/T_1$). Utilizing the MCMC algorithm, we derived the posterior distributions of these parameters. The light curves generated from these updated parameters show remarkable consistency with the original light curves. The fitting results are illustrated in Figure 9, and the corresponding parameters are listed in Table 2.

\begin{figure}[!ht]
\begin{center}
\begin{minipage}{5.7cm}
	\includegraphics[width=5.7cm]{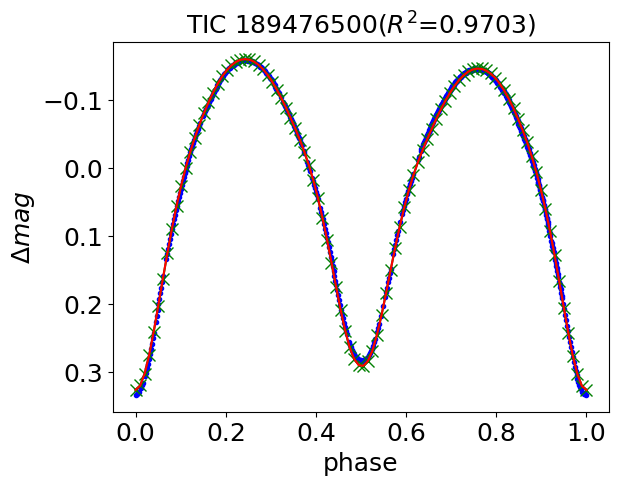}
\end{minipage}
\begin{minipage}{5.7cm}
	\includegraphics[width=5.7cm]{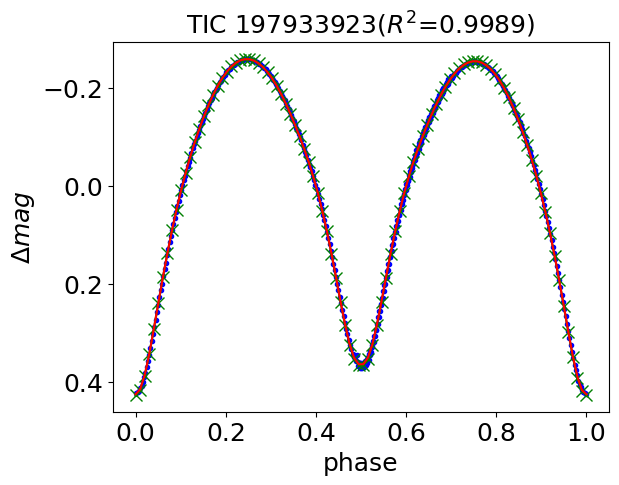}
\end{minipage}
\begin{minipage}{5.7cm}
	\includegraphics[width=5.7cm]{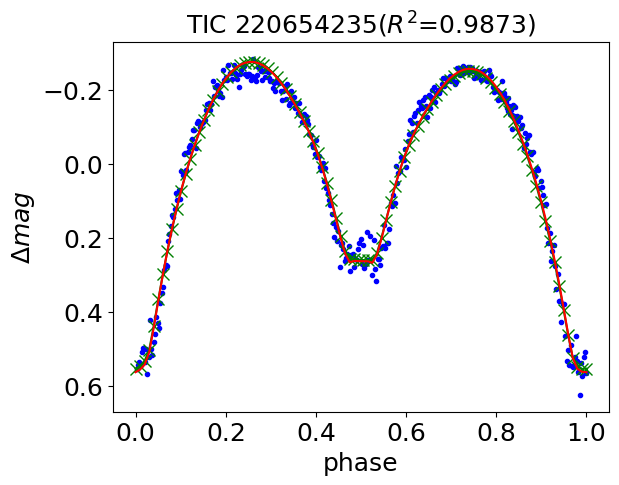}
\end{minipage}
\begin{minipage}{5.7cm}
	\includegraphics[width=5.7cm]{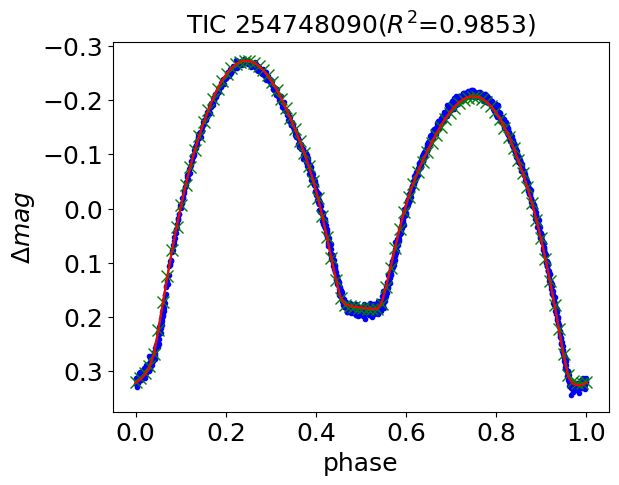}
\end{minipage}
\begin{minipage}{5.7cm}
	\includegraphics[width=5.7cm]{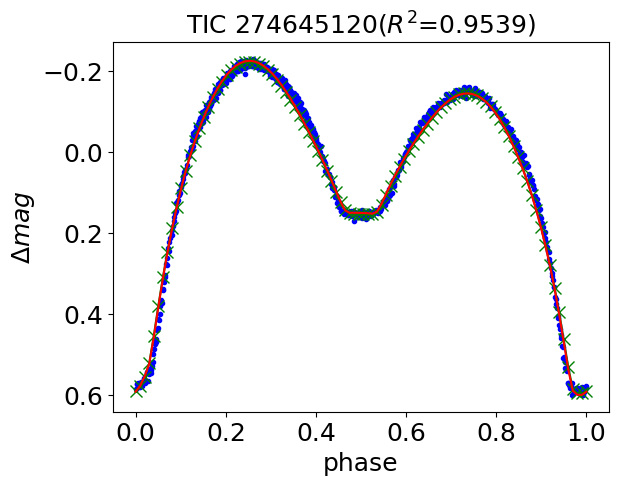}
\end{minipage}
\caption{The fitting results for five targets are depicted, with the blue dots indicating the original light curves. The red lines represent the light curves generated by the trained model (contact binary model), and the green 'x' dots denote the light curves produced by Phoebe.}\label{Fig8_1}
\end{center}
\end{figure}

\clearpage
{
\tiny
\begin{center}
\begin{longtable}{ccccccccccccccc}
\caption{The essential parameters of the five contact binary systems}\label{Table 2}\\
\hline\hline                          
name     &$T_1$(K)   &$incl(\circ)$  &$\sigma_{incl}$  &$q$  &$\sigma_{q}$  &$f$  &$\sigma_{f}$   &$\frac{T_2}{T_1}$  &$\sigma_{\frac{T_2}{T_1}}$  &$R^2$                   \\
\hline
\endhead
\hline

TIC 189476500 & 7178     & 76.228 & 0.477 & 0.298 & 0.016 & 0.320 & 0.007 & 0.957 & 0.016 & 0.970 \\
TIC 197933923 & 6559     & 80.926 & 0.234 & 0.425 & 0.007 & 0.688 & 0.014 & 0.994 & 0.002 & 0.999 \\
TIC 220654235 & 5639     & 86.856 & 2.506 & 0.356 & 0.015 & 0.791 & 0.103 & 0.970 & 0.016 & 0.987 \\
TIC 254748090 & 6347     & 87.597 & 1.679 & 0.243 & 0.005 & 0.914 & 0.038 & 0.967 & 0.005 & 0.985 \\
TIC 274645120 & 6265     & 89.010 & 4.563 & 0.338 & 0.038 & 0.898 & 0.291 & 0.793 & 0.065 & 0.954 \\

\hline
\end{longtable}
\end{center}
}

\subsection{Unknown types: 3 targets}
For the three targets - TIC 220521199, TIC 278075650, and TIC 463363407 - neither the more massive star filling the Roche lobe model nor the less massive star filling the Roche lobe model provided satisfactory fits, as demonstrated in Figure 10. There are two primary possible explanations for this discrepancy: first, these targets might belong to different types of variable stars, such as detached binaries; second, the parameter space of our sample may not fully encompass the parameters required for these three targets. Further in-depth analysis and research will be conducted to better understand these three targets.

\begin{figure}[!ht]
\begin{center}
\begin{minipage}{5.7cm}
	\includegraphics[width=5.7cm]{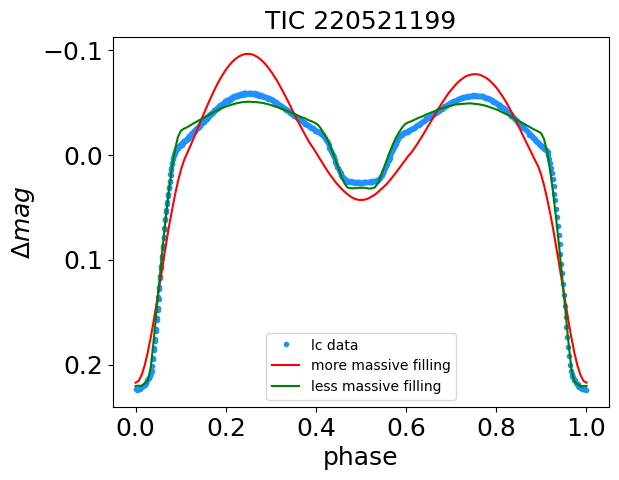}
\end{minipage}
\begin{minipage}{5.7cm}
	\includegraphics[width=5.7cm]{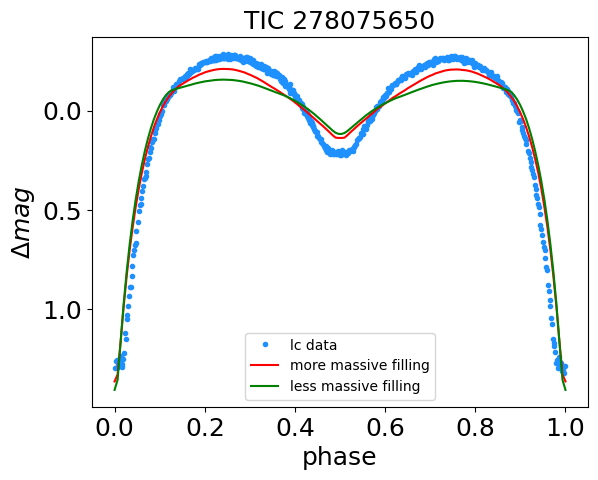}
\end{minipage}
\begin{minipage}{5.7cm}
	\includegraphics[width=5.7cm]{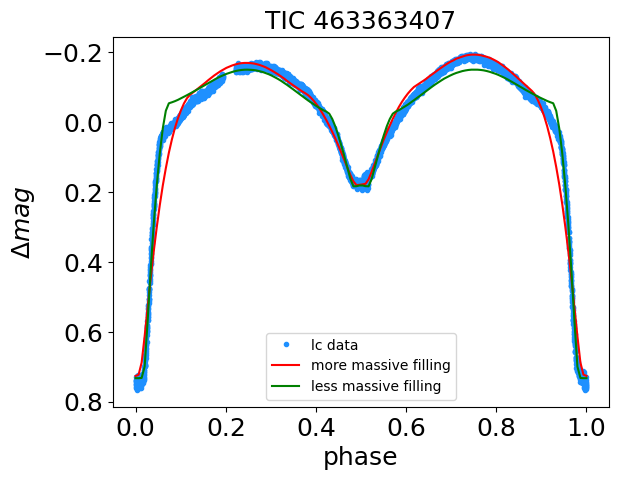}
\end{minipage}
\caption{The fitting results for three targets are depicted, with the light blue dots indicating the original light curves. The red lines represent the light curves generated by the trained model (more massive filling), and the green lines denote the light curves produced by the trained model (less massive filling).}\label{Fig89}
\end{center}
\end{figure}

\section{Discussion and Conclusion}
In these studies, we established a mapping relationship between the parameters of semi-detached binaries and their light curves. Our neural network model demonstrates a substantial computational advantage over the traditional Phoebe program in generating phase-magnitude light curves. Specifically, the model requires approximately 0.016 seconds to produce a light curve consisting of 150 data points on a system equipped with 32 GB of RAM and an i7-8700 CPU running at 3.20 GHz with 12 cores. In contrast, the Phoebe program takes roughly 5.488 seconds to generate a similarly sized light curve under identical hardware conditions. This comparison highlights that the neural network model is approximately two orders of magnitude faster than Phoebe, underscoring its efficiency and potential for enhancing the speed of astrophysical data analysis. By combining this with the Markov Chain Monte Carlo (MCMC) algorithm, we fitted the EB-type light curves published by \citet{Gao+et+al+2025}, resulting in the identification of 327 semi-detached binaries where the less massive star fills its Roche lobe, and three where the more massive star fills its Roche lobe. Notably, five targets could be well-fitted using models for contact binaries. Additionally, there were three targets that did not fit well, potentially due to two reasons, which we analyzed and discussed in Section 4.4. Future work will involve establishing detached binary models to perform detailed analyses of the three targets with poor fits, as well as attempting to further optimize the parameter space of the models developed in this study.

We intend to perform a cross-comparison with the currently recognized semi-detached binary star targets. Specifically, by matching the names sourced from the SIMBAD database, we aim to determine the number of targets that are absent from the known catalogs. In the known catalogs, numerous targets overlap one another. In the existing literature, \citet{Xiong+et+al+2024} have successfully identified 77 semi-detached binary candidates from TESS. \citet{Malkov+et+al+2020} compiled data to present a list of 119 semi-detached double-lined eclipsing binaries. In addition, \citet{Ibano+et+al+2006} listed 61 semi-detached close binaries, \citet{Surkova+et+al+2004} documented 232 semi-detached eclipsing binaries, and \citet{Lei+et+al+2024} reported eight semi-detached eclipsing binaries. Our cross-comparison analysis reveals that merely 26 targets in our catalog overlap with those in the previously mentioned catalogs. Consequently, it can be inferred that 304 targets are newly identified, which contributes to the expansion of the number of newly discovered semi-detached binary stars. To gain a deeper understanding of the reasons behind the emergence of such results, we conduct a more in-depth analysis. Specifically, we delve into the catalog presented by \citet{Xiong+et+al+2024}. Only 10 targets from our catalog are found in the one presented by \citet{Xiong+et+al+2024}. This study primarily conducts parameter inversion based on the EB-type light curves released by \citet{Gao+et+al+2025}. In contrast, many of the targets in the research of \citet{Xiong+et+al+2024} are not included in the EB-type catalog released by \citet{Gao+et+al+2025}, but instead belong to the EA-type and EW-type catalogs. 

We also provide the fundamental parameters of these targets. Numerical simulations performed by \citet{Terrell+et+al+2005} further substantiated the conclusion that photometric mass ratios for totally eclipsing semi-detached binaries are reliable. The catalogs from \citet{Surkova+et+al+2004} and \citet{Lei+et+al+2024} provide the parameters of $q$, $r_1$, $r_2$, and $incl$. The parameter catalog in this work has been compared with these two catalogs. Our results closely align with theirs. \citet{Surkova+et+al+2004} present a catalog that is divided into two sections: the first includes 96 widely recognized semi-detached eclipsing binaries with established photometric and spectroscopic data, and the second encompasses 136 semi-detached eclipsing binaries with documented photometric parameters yet lacking spectroscopic data. \citet{Lei+et+al+2024} performed a comprehensive photometric and spectroscopic investigation on eight such semi-detached eclipsing binaries. Within the catalog provided by \citet{Surkova+et+al+2004}, there are 11 targets included. Out of these, 6 are listed in the catalog that includes radial velocity parameters, while 5 are found in the catalog that lacks such information. Additionally, one target is identified in the catalog presented by \citet{Lei+et+al+2024}. The corresponding parameters derived from these targets are detailed in Table 3. In the paper by \citet{Lei+et+al+2024}, a total of 8 targets are identified, of which only one target is classified under the EB-type category in the work of \citet{Gao+et+al+2025}, whereas the remaining targets are categorized under the EA-type and ROT-type catalogs. Therefore, in the EA-type catalog obtained by \citet{Gao+et+al+2025}, there may be a significant number of semi-detached binaries, and we will conduct further research on this topic. From this perspective, it is evident that developing a neural network model for detached binaries to fit and disentangle EA-type light curves is highly necessary.

\begin{table*}
    \centering
	\caption{Parameter Comparison: $q, r_1, r_2, incl$ results from this study, $q^*, r_1^*, r_2^*, incl^*$ results from \citet{Surkova+et+al+2004} and \citet{Lei+et+al+2024}.}\label{Table 5}
	\begin{tabular}{ccccccccc} 
\hline \hline                        
 name &  $q$ &     $r_1$ &     $r_2$ &  $incl$ &   $q^*$ &      $r_1^*$ &      $r_2^*$ &   $incl^*$    \\
\hline
TIC $376162807^a$ & 0.41 & 0.22 & 0.30 & 89.92 & 0.41 & 0.20 & 0.30 & 88.70 \\
TIC $266770712^a$ & 0.57 & 0.42 & 0.33 & 86.93 & 0.51 & 0.41 & 0.32 & 87.00 \\
TIC $380117288^a$ & 0.69 & 0.35 & 0.35 & 84.05 & 0.61 & 0.37 & 0.33 & 80.80 \\
TIC $89761288^a$ & 0.60 & 0.18 & 0.34 & 89.92 & 0.56 & 0.13 & 0.33 & 83.20 \\
TIC $389916225^a$ & 0.49 & 0.44 & 0.32 & 89.84 & 0.44 & 0.44 & 0.30 & 87.70 \\
TIC $153991851^a$ & 0.54 & 0.42 & 0.33 & 76.70 & 0.53 & 0.43 & 0.32 & 74.20 \\
TIC $309178701^b$ & 0.54 & 0.39 & 0.33 & 82.01 & 0.59 & 0.38 & 0.34 & 82.00 \\
TIC $423387084^b$ & 0.30 & 0.47 & 0.28 & 82.60 & 0.26 & 0.48 & 0.26 & 83.70 \\
TIC $394881270^b$ & 0.51 & 0.19 & 0.32 & 87.47 & 0.51 & 0.17 & 0.32 & 79.30 \\
TIC $126602778^b$ & 0.52 & 0.37 & 0.32 & 75.20 & 0.45 & 0.36 & 0.30 & 73.60 \\
TIC $301347590^b$ & 0.42 & 0.19 & 0.31 & 79.51 & 0.40 & 0.19 & 0.29 & 80.00 \\	  
TIC $158431889^c$ & 0.14 &0.47	&0.22  &82.03  & 0.14  & 0.47  & 0.22  & 86.36 \\

\hline
\end{tabular}

\medskip
\footnotesize 
\textit{Note:} From the work of \citet{Surkova+et+al+2004}, $target^a$ is derived with photometric and spectroscopic data, while $target^b$ is derived with photometric data. $target^c$ is obtained from the work of \citet{Lei+et+al+2024}. \\
    
\end{table*}

\begin{acknowledgments}
We are very grateful for the data released by the TESS survey (https://archive.stsci.edu/missions-and-data/tess/). This work is supported by the Strategic Priority Research Programme of the Chinese Academy of Sciences (Grant No. XDB 41000000), the National Natural Science Foundation of China (Nos. 12103088 and 12433009), the National Key R\&D Program of China (Nos. 2022YFF0711500 and 2023YFA1608300), Yunnan Key Laboratory of Solar Physics and Space Science under No.202205AG070009, Yunnan Provincial Key Laboratory of Forensic Science  (No.YJXK005), Yunnan Basic Research Program (Nos. 202201AU070116 and 202501AT070027) and Yunnan Key Laboratory of Service Computing, Kunming, China. We acknowledge the science research grant from the China Manned Space Project under No.CMS-CSST-2021-A10, No.CMS-CSST-2021-A12 and No.CMS-CSST-2021-B10.
\end{acknowledgments}



\end{document}